\newcounter{myctr}

\documentclass{ws-ijqi}

\begin{document}

\markboth{Hans-Thomas Elze}
{The attractor and the quantum states}

\catchline{}{}{}{}{}

\title{THE ATTRACTOR AND THE QUANTUM STATES  
}

\author{HANS-THOMAS ELZE
}

\address{Dipartimento di Fisica, Universit\`a di Pisa,\\ Largo Pontecorvo 3, I--56127 Pisa, Italia
\\ elze@df.unipi.it}



\maketitle

\begin{history}
\received{\today}
\end{history}

\begin{abstract}
The dissipative dynamics anticipated in the proof of 't Hooft's existence theorem 
 -- {\it ``For any quantum system there exists at least one deterministic model that 
reproduces all its dynamics after prequantization''} -- 
is constructed here explicitly. We propose a generalization of 
Liouville's classical phase space equation, incorporating dissipation and diffusion, and 
demonstrate that it describes the emergence of quantum states
and their dynamics in the Schr\"odinger picture. Asymptotically, 
there is a stable ground state and two decoupled sets of degrees of freedom, which 
transform into each other under the energy-parity 
symmetry of Kaplan and Sundrum. They recover the familiar Hilbert space and its dual. 
Expectations of observables are shown 
to agree with the Born rule, which is not imposed {\it a priori}.   
This attractor mechanism is applicable in the presence of interactions,    
to few-body or field theories in particular. 

\end{abstract}


\section{Introduction}	
There is a growing number of deterministic models of quantum mechanical objects 
which are based on conjectured fundamental information loss or dissipation 
mechanisms~\cite{tHooft06,Elze05,Blasone05,Adler,Smolin,Vitiello01}. 

These studies are largely motivated by  
the unresolved issues surrounding ``quantum gravity'', i.e., by the conflict 
between quantum mechanics necessitating an external time and 
diffeomorphism invariance in general relativity, for example, which defies 
its existence. Furthermore, despite its great successes in describing 
the statistical aspects of experiments, quantum theory itself presents  
problems of interpretation, which arise from its indeterministic features 
and which are clearly seen, for example, in the unresolved measurement problem (also 
``wave function collapse'' or ``objective reduction''). Thus,      
concerning the foundations of quantum mechanics, there is an increasing impetus 
to try to reconstruct and to better understand the emergence of quantum 
mechanics from simpler structures beneath.   

So far, the construction of models has proceeded case by case.  
It is guided by the idea that quantum states may actually represent 
large equivalence classes of deterministically evolving ``classical'' states --   
which become indistinguishable when affected by the conjectured 
information loss or dissipative process~\cite{tHooft06}.  

Furthermore, 't Hooft's existence theorem~\cite{tHooft07} shows that generally 
the evolution of all quantum mechanical objects that are characterized by a finite 
dimensional Hilbert space can be captured by a dissipative process. This has recently been 
generalized for objects which are described by a set 
of mutually commuting Hermitian operators~\cite{Elze08}, 
i.e., for the case of a finite set of {\it beables}~\cite{BellBook}. 
For completeness, we review these results in the Appendix. 
  
However, it has not been shown before how to complement the existence theorem   
by a dynamical theory. In order to describe the quantum mechanical world 
around us, such a theory has to deal with interacting as well as
approximately isolated objects that exist throughout a large 
range of length and energy scales.  
We make a step in this direction, by showing 
that familiar aspects of quantum mechanics can be generated by an  
attractor mechanism, which is obtained by a  
generalization of the classical Liouville equation.      
  
This will lead us to emerging quantum states which evolve according to the 
Schr\"odinger picture embodied in  
the von\,Neumann equation, to the Born rule, and to the Randall-Sundrum energy-parity 
symmetry, which may protect the cosmological constant against far too large corrections, 
which otherwise should be determined by particle physics scales~\cite{KS,Elze07}.

\section{A useful reformulation of Hamiltonian dynamics}

For simplicity, we consider objects 
with a single continuous degree of freedom. However, it is straightforward to 
repeat the following derivations for interacting few-body systems 
and fields. 
  
To begin with, we assume that there are only conservative forces and that Hamilton's equations 
are determined by the generic Hamiltonian function:  
\begin{equation}\label{HamiltonianF} 
H(x,p):=\frac{1}{2}p^2+V(x) 
\;\;, \end{equation} 
defined in terms of generalized coordinate $x$ and momentum $p$, and where  
$V(x)$ denotes the potential. -- An ensemble of such objects, for example, 
following trajectories with different initial conditions, is described by 
a distribution function $f$ in phase space, i.e., by the probability 
$f(x,p;t)\mbox{d}x\mbox{d}p$ to find a member of the ensemble in an infinitesimal 
volume at point $(x,p)$. This distribution evolves according to the Liouville equation: 
\begin{equation}\label{LiouvilleEq} 
-\partial_tf=\frac{\partial H}{\partial p}\cdot\frac{\partial f}{\partial_x}
-\frac{\partial H}{\partial x}\cdot\frac{\partial f}{\partial_p}
=\big\{ p\partial_x-V'(x)\partial_p\big\}f 
\;\;, \end{equation} 
with $V'(x):=\mbox{d}V(x)/\mbox{d}x$. 

A Fourier transformation, $f(x,p;t)=\int\mbox{d}y\;\mbox{e}^{-ipy}f(x,y;t)$, 
replaces the Liouville equation by: 
\begin{equation}\label{LFourier}  
i\partial_tf=\big\{ -\partial_y\partial_x+yV'(x)\big\}f 
\;\;, \end{equation} 
without changing the symbol for the distribution function,  
whenever changing variables.  
Thus, momentum is eliminated in favour of {\it doubling} the number of coordinates. Finally, 
with the transformation: 
\begin{equation}\label{coordtrans} 
Q:=x+y/2\;\;,\;\;\;q:=x-y/2  
\;\;, \end{equation} 
we obtain the Liouville equation in the form: 
\begin{eqnarray}\label{Schroed} 
i\partial_tf&=&\big\{ \hat H_Q-\hat H_q+\Delta (Q,q)\big\}f
\;\;, \\ [1ex] \label{HX} 
\hat H_\chi &:=&-\frac{1}{2}\partial_\chi ^{\;2}+V(\chi )\;\;, 
\;\;\;\mbox{for}\;\;\chi =Q,q 
\;\;, \\ [1ex] \label{I} 
\Delta (Q,q)&:=&(Q-q)V'(\frac{Q+q}{2})
-V(Q)+V(q)\;=\;-\Delta (q,Q)
\;\;. \end{eqnarray}  
Several comments are in order here: 
\begin{itemize} 
\item The presented reformulation of classical dynamics 
is rather independent of the number of degrees of freedom. 
It applies to matrix valued as well as to   
Grassmann valued variables, representing the ``pseudoclassical'' fermion  
fields introduced by Casalbuoni and by Berezin and Marinov. 
Field theories require the classical 
functional formalism employed previously in a related context~\cite{Elze05,Elze07}. 
Gauge theories or, generally, theories with constraints 
have to be examined carefully.  \\ \noindent  
\item The Eq.\,(\ref{Schroed}) appears as the {\it von\,Neumann equation} 
for a density operator $\hat f(t)$, 
considering $f(Q,q;t)$ as its matrix elements. However, a crucial difference is found  
in the interaction $\Delta$ between the {\it bra-} and {\it ket-} states.  
The related Hilbert space and its dual, therefore are coupled, unlike  
the case of quantum mechanics. \\ \noindent     
\item Alternatively, the Eq.\,(\ref{Schroed}) might be read as the     
{\it Schr\"odinger equation} for two identical (sets of) degrees of freedom. 
However, their respective Hamilton operators, $\hat H_{Q,q}$, contribute with opposite sign. 
Since their interaction $\Delta$ is antisymmetric under 
$Q\leftrightarrow q$, the complete   
(Liouville) operator on the right-hand side of Eq.\,(\ref{Schroed}) has a symmetric spectrum 
with respect to zero and, generically, will not be bounded below. 
This {\it Kaplan-Sundrum energy parity symmetry} has been invoked before as a protection  
for a (near) zero cosmological constant which, otherwise, is threatened by 
many orders of magnitude too large zeropoint energies~\cite{KS,Elze07}.  
\end{itemize} 

Only for free particle or harmonic oscillator $\Delta$ vanishes. -- Generally, with a coupling 
of the Hilbert space and its dual, or without a stable ground state, 
our reformulation of Hamiltonian dynamics does not qualify as a quantum theory.  

The problem of a missing ground state, sometimes in disguise, has been the stumbling block 
in previous attempts at deterministic model building for quantum objects and has been overcome 
only in individual examples~\cite{tHooft06,Elze05,Blasone05,Adler,Smolin,Vitiello01}. 

\section{An attractor mechanism and emergent quantum features}

Based on Hamilton's equations, we have seen that deterministic ensemble theories 
imply the absence of a lowest energy state, when suitably rewritten in the form 
of a Schr\"odinger equation (or functional Schr\"odinger equation for field theories).  
On the other hand, a dissipative process was of paramount importance in 
the proof of existence of deterministic models for quantum mechanical 
objects, as reviewed in the Appendix. -- In the following, we shall make use of these hints. 

We propose an attractor mechanism that turns    
deterministic evolution, described by an ensemble theory as in 
Section\,2., into the Schr\"odinger evolution of 
quantum states, based on a stable ground state.  
Our construction refers particularly to two assumptions: \\ \noindent 
(A) {\it The emergence of quantum states originates from a microscopic  
process beneath which applies to all physical objects.} \\ \noindent   
(B) {\it The statistical interpretation of quantum states (Born rule) 
originates from the classical ensemble theory.}  

\subsection{Expectations, operators and the Born rule}

We begin with the normalization of the classical probability distribution: 
\begin{equation}\label{clnorm} 
1\stackrel{!}{=}\int\frac{\mbox{d}x\mbox{d}p}{2\pi}\;f(x,p;t)=\int\mbox{d}Q\mbox{d}q\;
\delta (Q-q)f(Q,q;t)=:\mbox{Tr}\;\hat f(t) 
\;\;, \end{equation}  
incorporating the transformations of Section\,2. Consider a    
complete set of orthonormal eigenfunctions of the operator $\hat H_\chi $ of Eq.\,(\ref{HX}), 
defined by $g_j(\chi ;t):=\mbox{exp}(-iE_jt)g_j(\chi )$ and 
$\hat H_\chi g_j(\chi )=E_jg_j(\chi )$, respectively, 
with a discrete spectrum, for simplicity. 
Then, we may expand $f$: 
\begin{equation}\label{fexpans} 
f(Q,q;t)=\sum_{j,k}f_{jk}(t)g_j(Q;t)g_k^*(q;t)
\;\;. \end{equation} 
Employing this, the normalization condition (\ref{clnorm}) can be stated as: 
\begin{equation}\label{clnorm1} 
1\stackrel{!}{=}\sum_{j,k}f_{jk}(t)\mbox{e}^{-i(E_j-E_k)t}\int\mbox{d}Q\;g_j(Q)g_k(Q)
=\sum_{j}f_{jj}(t)
\;\;. \end{equation} 
Since the classical phase space 
distribution is real, the expansion coefficients form 
a Hermitean matrix, $f_{ij}=f^\ast_{ji}$, which we also denote by $\hat f$.  
  
The {\it classical} expectation values are calculated as follows: 
\begin{eqnarray}\label{xexpect} 
\langle x\rangle :=\int\frac{\mbox{d}x\mbox{d}p}{2\pi}\;xf(x,p;t)
&=&\int\mbox{d}Q\mbox{d}q\;\delta (Q-q)\frac{Q+q}{2}f(Q,q;t)
\;\;, \\ [1ex] \label{xexpect1} 
&=:&\mbox{Tr}\;\big (\hat X\hat f(t)\big ) 
\;\;, \\ [1ex] \label{pexpect} 
\langle p\rangle :=\int\frac{\mbox{d}x\mbox{d}p}{2\pi}\;pf(x,p;t)
&=&\int\mbox{d}Q\mbox{d}q\;\delta (Q-q)(-i)\frac{\partial_Q-\partial_q}{2}f(Q,q;t)
\;\;, \\ [1ex] \label{pexpect1} 
&=:&\mbox{Tr}\;\big (\hat P\hat f(t)\big ) 
\;\;, \end{eqnarray}  
introducing the operators $\hat X$ and $\hat P$, with matrix elements 
$X(q,Q)=\delta (Q-q)(Q+q)/2$ and 
$P(q,Q)=-i\big (\delta (Q-q)\stackrel{\rightharpoondown}{\partial}_Q-
\stackrel{\leftharpoondown}{\partial}_q\delta (Q-q)\big )$ (derivatives act left 
or right, as indicated). 
Eliminating one of the two integrations in the above equations with the help 
of the $\delta$-functions and suitable partial integrations, these operators are 
recognized as the coordinate and momentum operators of quantum theory. 

We also find, for example: 
\begin{equation}\label{noncommops} 
\mbox{Tr}\;\big ((\hat X\hat P+\hat P\hat X)\hat f(t)\big ) = 
\frac{i}{2}+2\int\frac{\mbox{d}x\mbox{d}p}{2\pi}\;xpf(x,p;t)-\frac{i}{2} 
\;\;, \end{equation} 
where the cancelling imaginary terms stem from the first and second 
term on the left-hand side, respectively, which both contribute equally to the 
integral. 
 
Operators appear here strictly by rewriting classical  
statistical formulae and {\it not by a quantization rule}, such as 
replacing $x$ and $p$ by operators $\hat X$ and $\hat P$, with $[\hat X,\hat P]=i$, 
acting on a Hilbert space (not necessarily related to phase space).   

Furthermore, the Eqs.\,(\ref{clnorm}), (\ref{xexpect})--(\ref{noncommops}) are in accordance 
with the interpretation of $f(Q,q;t)$ as matrix elements of a density operator $\hat f(t)$. 

However, there is an important {\it caveat}: The eigenvalues of normalized 
quantum mechanical density operators are usually constrained to lie between zero and one, 
corresponding to the interpretation as standard probabilities.  
This is not necessarily the case with the 
operator $\hat f$ obtained from a classical probability distribution. Similarly, 
the Wigner distribution -- obtained from the matrix elements of 
a quantum mechanical density operator by applying the   
transformations leading from $f(x,p)$ to $f(Q,q)$ in reverse -- generally, is not 
positive semi-definite on phase space, even though its marginal distributions are. 
Therefore, it does not necessarily qualify as a  
classical probability density.  
  
The discussion of negative or larger-than-one probabilities is beyond the scope 
of the present article; we refer to the literature for arguments that make 
sense of them, see, for example, Refs.~\cite{Dirac,Feynman,Khrennikov,Hartle}. 
However, we anticipate that the proper quantum mechanical 
aspects of the states shall emerge dynamically, see Section\,3.3. We will point out in the 
following, where the restriction to probabilities that lie in the interval $[0,1]$ 
arises. Thus, 
the application of operators here, together with the density operator in 
particular, constitute examples for the {\it Born rule}, once they are complemented 
by a suitable dynamical framework. 

\subsection{Spacetime fluctuations and dissipative dynamics}

We will propose a generalization of the conservative classical dynamics 
described by Eqs.\,(\ref{Schroed})--(\ref{I}) that incorporates dissipation 
as well as diffusion. For sufficiently long times,   
the evolving density operator should be attracted to solutions 
of the quantum mechanical von\,Neumann equation: 
\begin{equation}\label{vonNeumann} 
i\partial_t\hat f=[\hat H_\chi ,\hat f] 
\;\;. \end{equation}   
Equivalently, the expansion coefficients $f_{ij}$ in Eq.\,(\ref{fexpans}) 
should become constants.   

Employing this expansion, generally, the Eq.\,(\ref{Schroed}) can be 
rewritten as matrix equation for the coefficients: 
\begin{equation}\label{Schroedcoeff}   
i\partial_tf_{jk}(t)=\sum_{l,m}\Delta_{jklm}f_{lm}(t) 
\;\;, \end{equation} 
with the interaction double matrix $\hat\Delta$ defined by: 
\begin{equation}\label{Deltamatrix}  
\Delta_{jklm}:=\int\mbox{d}Q\mbox{d}q\;g_j(Q)g_k(q)\Delta (Q,q)g_l(Q)g_m(q) 
=-\Delta_{kjml} 
\;\;,  \end{equation} 
employing the antisymmetry of $\Delta$, Eq.\,(\ref{I}). 
Consequently, $i\Delta$ maps a Hermitean matrix, such as $\hat f$, 
to a Hermitean matrix. This map produces zero  
when taking the trace, $\mbox{Tr}\;(\hat\Delta\hat M)=0$, 
for any matrix $\hat M$, since: 
\begin{equation}\label{TrZero} 
\sum_j\Delta_{jjlm}=0 
\;\;, \end{equation} 
by definition (\ref{Deltamatrix}), completeness, and Eq.\,(\ref{I}). 
For example, the solution of Eq.\,(\ref{Schroedcoeff}), 
$\hat f(t)=\mbox{exp}(-i\hat\Delta t)\hat f(0)$,  
conserves the normalization of $\hat f$, Eq.\,(\ref{clnorm1}).  

The interaction $\Delta$   
presents the unfamiliar coupling between 
Hilbert space and its dual, 
which prevented us from considering 
Eq.\,(\ref{Schroed}) as a truly quantum mechanical equation. We will now  
present some heuristic considerations which entail important   
modifications of the deterministic ensemble theory.

For very small intervals, spacetime itself may be thought to have   
atomistic structure and dynamics, as discussed, for example, in terms of 
classical {\it causal sets}.  
Such a locally finite ordered set evolves by sequential growth, 
i.e., by the random (``sprinkling'') appearance of new set 
elements~\cite{SorkinRev,Dowker,RideoutSorkin1}. 

Consider a ``hypersurface'' formed by the set elements that have no 
successors -- i.e., which do not precede any other set elements, according to 
the causal order relation -- at a certain stage of the evolution. 
Furthermore, consider the probability distribution $f$ of our coarse grained 
phase space picture and, in particular, the amount of probability 
that resides in the volume $V\propto N$ corresponding to $N$ arbitrarily chosen 
elements in this hypersurface. Following further growth, with a 
sufficiently large number of new elements added to the causal set, we pick   
those $N+\delta N$ elements from the evolved hypersurface that 
either have ancestors among the previously chosen $N$ elements or that are among 
the previously chosen elements and have not become ancestor to any new 
element. If $\delta N\neq 0$, this induces a change of probability density, if 
the amount of probability is conserved, as the volume evolves correspondingly, 
$V\rightarrow V+\delta V\propto N+\delta N$. 

Comparing two different choices of the $N$ elements, say subsets $A$ and $B$ 
of the considered hypersurface, the corresponding values $\delta N_A$ and $\delta N_B$, 
generally, will differ, due to the random growth of the causal set. 
This induces random fluctuations in the probability density.     

We represent the induced fluctuations by a homogeneous stochastic term $i\delta Hf$ contributing to the right-hand side of Eq.\,(\ref{LiouvilleEq}). We shall see shortly that it leads 
to {\it dissipation}.  

Relatively, such fluctuations become less important, i.e.,   
typical fluctuations decrease, as the discrete spacetime associated with the 
growing causal set continues to evolve, supposedly towards a continuum limit. 
Accordingly, we treat $\delta H$ as a random variable, with 
distribution $\propto\exp (-t\delta H^2/4\epsilon)$ and $\epsilon$ an energy scale. 
Variability (``aging'') of $\epsilon$, incorporating an initial cut-off of the width, 
for example, is neglected.       

Furthermore, the evolution of $f$ must be modified by {\it diffusion}, caused by  
the randomly growing structure beneath~\cite{DowkerHensonSorkin04}. 
Somehow reminding of {\it Galton's board}, additional causal set elements keep appearing  
on which a moving phase space distribution ``scatters'' and, thus, spreads   
into different microscopic evolutionary paths. 

Asymptotically, diffusion and dissipation may balance each other in 
such a way that probability is conserved, resulting in 
a Hermitian matrix $\hat g$, with unit trace, to which $\hat f$ 
must be attracted.  

These effects are incorporated in our minimalist model: 
\begin{equation}\label{Schroedcoeff1}   
i\partial_t\hat f(t)=(\hat\Delta +\delta H)(\hat f(t)-\hat g(t)) 
\;\;, \end{equation} 
which generalizes Eq.\,(\ref{Schroedcoeff}).    
The dissipative term stems from $+\delta Hf(Q,q)$ entering 
the right-hand side of Eq.\,(\ref{Schroed}), which means adding   
$+\delta H\sum_{l,m}\delta_{jl}\delta_{km}f_{lm}(t)=\delta Hf_{jk}(t)$ in 
Eq.\,(\ref{Schroedcoeff}).  
Together with its prefactors, the matrix $\hat g$ enters as a source term here. 

The solution of the linear first order Eq.\,(\ref{Schroedcoeff1}) is: 
\begin{equation}\label{fdisssol} 
\hat f(t)=\mbox{e}^{-i(\hat\Delta +\delta H)t}\Big (\hat f(0)+i(\hat\Delta +\delta H)
\int_0^t\mbox{d}s\;\mbox{e}^{i(\hat\Delta +\delta H)s}\hat g(s)\Big )
\;\;. \end{equation} 
Averaging over the Gaussian fluctuations $\delta H$ gives: 
\begin{eqnarray}
\hat f(t)&=&\mbox{e}^{-i(\hat\Delta -i\epsilon )t}\Big (\hat f(0)+i(\hat\Delta -i\epsilon )
\int_0^t\mbox{d}s\;\mbox{e}^{i(\hat\Delta -i\epsilon )s}\hat g(s)\Big )
\nonumber \\ [1ex] \label{fdisssol1} 
&=&\hat g(t)+\mbox{e}^{-i(\hat\Delta -i\epsilon )t}\big (\hat f(0)-\hat g(0)\big ) 
-\int_0^t\mbox{d}s\;\mbox{e}^{-i(\hat\Delta -i\epsilon )(t-s)}\partial_s\hat g(s)
\;\;, \end{eqnarray} 
which shows the nonunitary dissipative decay caused by the fluctuations. 
Taking the trace of 
Eq.\,(\ref{fdisssol}) or (\ref{fdisssol1}), with the help of Eq.\,(\ref{TrZero}) and 
with $\mbox{Tr}\;\hat f(0)=1$, however, we find that probability is 
conserved, i.e., the normalization $\mbox{Tr}\;\hat f(t)=1$, 
provided that $\mbox{Tr}\;\hat g(t)=1$. 
Thus, the source term compensates the dissipative loss.  
  
Furthermore, if $\hat g(t)$ becomes constant sufficiently fast, for large 
$t\gg 1/\epsilon$, then $\hat f(t)\approx\hat g(t)\rightarrow\hat g(\infty )$, 
asymptotically. In this limit, dissipation effectively eliminates the 
coupling $\Delta$ and our simplistic account of dissipation/diffusion leads to constant 
matrix elements $f_{ij}(t)\rightarrow g_{ij}(\infty )$. -- Via Eq.\,(\ref{fexpans}), 
this implies that the von\,Neumann Eq.\,(\ref{vonNeumann}) becomes a valid 
approximation for sufficiently large $t$; i.e., in this limit      
{\it quantum theory} will be recovered. 

However, we still need a relation between 
the matrices $\hat g(t)$ and $\hat f(t)$, in order to complete the dynamical 
model and to take into account that the properties of common quantum mechanical 
objects are highly reproducible. 

\subsection{Attraction towards quantum states}

The minimalist model of Eq.\,(\ref{Schroedcoeff1}) is completed by a 
nonlinear equation for the source matrix $\hat g$: 
\begin{equation}\label{nonling}
\partial_t\hat g(t)=\tau^{-1}
\big (\hat f(0)-\langle\hat f(0)\rangle_{\hat g(t)}\big )\hat g(t) 
\;\;, \end{equation} 
with $\tau$ a time scale and  
$\langle\hat f\rangle_{\hat g}:=\mbox{Tr}\;(\hat f\hat g)/\mbox{Tr}\;\hat g$. 

This equation is structurally similar to nonlinear wave equations that 
have been considered as extensions of the Schr\"odinger equation, incorporating 
dissipation or measurement processes into quantum 
mechanics~\cite{Gisin81,Grigorenko95,GisinRigo95}. However, presently we consider it  
as a phenomenological description of how -- following a measurement or state preparation --   
the source evolves. The diffusive effects, alluded to before, 
are assumed to result in a source term which reflects the 
initial condition of the classical theory, represented by 
$\hat f(0)$. 
  
The solution of Eq.\,(\ref{nonling}) is given simply by: 
\begin{equation}\label{nonlingsol} 
\hat g(t)=\mbox{e}^{\hat f(0)t/\tau}/\mbox{Tr}\;\mbox{e}^{\hat f(0)t/\tau} 
\;\;. \end{equation}   
This implies $\mbox{Tr}\;\hat g(t)=1$, as needed for the conservation of probability. 

For all times, the matrix $\hat g(t)$ can be diagonalized by the unitary transformation 
which diagonalizes the Hermitian matrix $\hat f(0)$, 
$\hat U\hat f(0)\hat U^{\dagger}=
\mbox{diag}(f'_{11}(0),\;f'_{22}(0),\;\dots\; )$.
Furthermore, for {\it all} finite initial $\hat f(0)$, 
the eigenvalues of the matrix $\hat g(t)$ are positive and, 
by its unit trace, constrained to lie in $[0,1]$. In particular, note that    
initially $\hat g(0)=\hat{\mathbf{1}}/\mbox{Tr}\;\hat{\mathbf{1}}$, i.e., 
the matrix starts out with a homogeneous distribution of eigenvalues 
(with a regularization to be implemented). This gives no preference to any energy eigenstate, 
recalling that we are working with the basis of eigenstates of the operator $\hat H_\chi $, 
cf. Section\,3.1.  

However, for sufficiently large times, the diagonalized matrix 
$\hat g_{\mbox{d}}(t):=\hat U\hat g(t)\hat U^{\dagger}$ approaches 
exponentially fast an onedimensional {\it projector} $\hat P$: 
\begin{eqnarray}\label{projector}
\hat g_{\mbox{d}}(t)&=&\frac{1}{\sum_j\;\mbox{e}^{(f'_{jj}-\lambda )t/\tau}} 
\mbox{diag}\left (\mbox{e}^{(f'_{11}-\lambda )t/\tau},\;\dots ,\; 
\mbox{e}^{(f'_{kk}-\lambda )t/\tau},\;\dots\; \right )
\\ [1ex] \label{projector1} 
&\longrightarrow&\;\;\mbox{diag}\left (0,\;\dots ,1,0,\;\dots\; \right )\;=:\hat P
\;\;, \end{eqnarray}
where the only novanishing 
entry appears in the position of the largest eigenvalue of $\hat g_{\mbox{d}}$, 
$\lambda :=\mbox{max}_k\;f'_{kk}$, assuming that it is not degenerate, for simplicity. 
This particular position is indicated by $\bar l$, for example, in  
$P_{jk}=\delta_{j\bar l}\delta_{k\bar l}$. 
 
We recall from the previous Section\,3.2. that 
the time dependent matrix $\hat f$ of expansion coefficients of Eq.\,(\ref{fexpans}) 
is attracted to the asymptotic value of the source matrix $\hat g$, 
for sufficiently large $t$. Undoing the diagonalizing transformation, 
which depends on $\hat f(0)$, 
we obtain: 
\begin{equation}\label{asympP}
\hat f(t)\;\longrightarrow\;\hat g(\infty )=\hat U^{\dagger}\hat P\hat U 
\;\;. \end{equation}   

Thus, if $\hat f(0)$ is diagonal, corresponding to $\hat U=\hat U^{\dagger}=\hat{\mathbf{1}}$, 
then the distribution $f$ of Eq.\,(\ref{fexpans}) is attracted to become a 
{\it density matrix representing a stationary state}:  
\begin{equation}\label{statstate} 
f(Q,q;t)\;\longrightarrow\;\sum_{j,k}P_{jk}g_j(Q;t)g^*_k(q;t)
=g_{\bar l}(Q)g_{\bar l}(q)
\;\;, \end{equation} 
which presents a solution of the von\,Neumann equation (\ref{vonNeumann}) 
with $\partial_t\hat f=0$. --   
All diagonal $\hat f(0)$ that have their largest 
eigenvalue in the same position (refering to the chosen basis), 
and which are related to classical phase space 
distributions via Eq.\,(\ref{fexpans}), lead to the same density matrix. In this 
precise sense, a stationary {\it quantum state represents a large 
equivalence class of classical distributions}.    

More generally, if $\hat f(0)$ is not diagonal, corresponding to 
$\hat U\neq\hat{\mathbf{1}}$, 
then the distribution $f$ is attracted to become a {\it density matrix representing a 
pure state}:  
\begin{equation}\label{purestate} 
f(Q,q;t)\;\longrightarrow\;\sum_{j,k}(\hat U^{\dagger}\hat P\hat U)_{jk}
g_j(Q;t)g^*_k(q;t)
\;\;. \end{equation} 
Also a solution of Eq.\,(\ref{vonNeumann}), this   
includes superpositions 
of stationary states, which are again seen as representing large 
equivalence classes of classical distributions. -- Finally, 
mixed states are formed by ensembles of pure states. 
  
Thus, the quantum states emerge asymtotically, exponentially fast. They are represented by  
properly normalized density matrices with eigenvalues in $[0,1]$, which  
agrees with the standard probability interpretation. 

However, this does not exclude initial phase 
space distributions $f$ which assume nonstandard values according to the underlying 
ensemble theory, cf. Section\,3.1. -- Here we share a 
pragmatic point of view: Due to the 
normalization of a phase space distribution $f$, any negative or larger-than-one 
probability must be compensated so that the sum over all alternatives 
of events, to which probabilities are assigned, equals one. Therefore, it must 
be possible to perform a coarse graining, i.e., a re-partitioning of the 
set of alternatives, such that nonstandard probabilities are avoided altogether. 
This amounts to a partitioning of the space of events, e.g., phase space, corresponding 
to independent alternatives which, in principle, are the ones that can be 
explored experimentally~\cite{Feynman}.     

In order to address such issues in detail,  
a microscopic theory is needed of dissipation and diffusion, 
due to spacetime fluctuations, and an  
understanding of how a measurement or state preparation fixes an initial 
$\hat f(0)$ or an equivalence class of such initial state matrices. 
This may lead to new insights concerning the measurement problem. Its solution  
must transcend quantum theory, in its usual formulation, as is well known. In particular, the 
stochastic nature of measurement results might be related to the fact that quantum states 
represent large equivalence classes of classical phase space distributions, as we 
have seen, rather than ontological ``elements of reality''.       


\section{Conclusions}
We have generalized Liouville's phase space equation for 
Hamiltonian systems by incorporating an attractor mechanism, comprising a 
homogeneous stochastic term and a source term. The latter compensates 
dynamically the dissipative loss of 
information induced by the former. -- For suitably chosen variables, 
this classical model can be cast into the form of Eqs.\,(\ref{Schroedcoeff1}) 
and (\ref{nonling}).  

Its solutions are attracted to 
density matrices which solve the von\,Neumann equation. 
With respect to the classical ensemble theory, 
these quantum states are recognized as large equivalence classes 
of states with varying initial conditions.  

Motivated by the assumption of an atomistic structure of spacetime 
-- such as represented by a causal set -- our model reflects        
the question: Does quantum mechanics originate, as a phenomenon of coarse graining, 
from fluctuations which are induced by the growth of discrete spacetime beneath? 

\section*{Acknowledgments}

It is a pleasure to thank A.\,Khrennikov and V. Manko for discussions 
and M.\,Genovese for the invitation 
to present this work at the 4th Workshop ad memoriam of 
Carlo Novero {\it ``Advances in Foundations of Quantum Mechanics and Quantum Information 
with Atoms and Photons''} (Torino, May 2008).  

\section*{Appendix: Deterministic models of quantum objects do exist}

Having mentioned individual cases of deterministic dynamical models of  
quantum objects in Section\,1, we review here statements 
about their existence in general.   

The existence theorem of 't\,Hooft concerns the Schr\"odinger equation 
for a quantum system with a $d$-dimensional Hilbert space~\cite{tHooft07}: 
\begin{equation}\label{G1} 
\frac{\mbox{d}\psi}{\mbox{d}t}=-i\hat H\psi  
\;\;, \end{equation} 
where $\hat H$ denotes the Hamiltonian, a $d\times d$ 
matrix here. 

As it turns out, the dynamics of Eq.\,(\ref{G1}) is reproduced 
in a {\it deterministic system} with two degrees of freedom, one periodic variable, 
$\varphi\in [0,2\pi [$, and another real variable, $\omega$, which evolve 
according to the classical equations of motion:    
\begin{eqnarray}\label{G2}
&&\frac{\mbox{d}\varphi (t)}{\mbox{d}t}=\omega 
\;\;, \\ [1ex] \label{G3}
&&\frac{\mbox{d}\omega (t)}{\mbox{d}t}=-\kappa f(\omega )f'(\omega )\;\;,\;\;\; 
f(\omega ):=\det\;(\hat H-\omega )
\;\;, \end{eqnarray}  
where $\kappa >0$ is a parameter. -- It is easy to see 
that $\omega$ moves exponentially 
fast towards one of the eigenvalues of $\hat H$, since multiplying $f$ by minus one
times its derivative $f'$ makes all corresponding zeros 
attractive (see Figure\,1 of Ref.\,\refcite{tHooft07} for an illustration).  
The initial condition for 
Eq.\,(\ref{G3}) determines which eigenvalue $E_i$ is approached, resulting    
in a limit cycle for $\varphi$ with period 
$T_i\equiv 2\pi\omega_i^{-1}=2\pi E_i^{-1}$. 

It is useful to introduce two auxiliary operators: 
\begin{equation}\label{momenta} 
\hat p_\varphi :=-i\frac{\partial}{\partial\varphi}
\;\;,\;\;\;  
\hat p_\omega :=-i\frac{\partial}{\partial\omega}
\;\;, \end{equation} 
which are {\it not} related to classical observables. 
We also define: 
\begin{equation}\label{evolutionop} 
\hat h:=\omega\hat p_\varphi -\frac{\kappa}{2}
\{ f(\omega )f'(\omega ),\hat p_\omega \}
\;\;, \end{equation} 
with $\{ x,y\}:=xy+yx$. This operator generates the evolution described by the classical 
equations of motion (\ref{G2})--(\ref{G3}). Indeed, we can rewrite them as: 
\begin{eqnarray}\label{G21}
&&\frac{\mbox{d}\varphi (t)}{\mbox{d}t}=-i[\varphi (t),\hat h] 
\;\;, \\ [1ex] \label{G31}
&&\frac{\mbox{d}\omega (t)}{\mbox{d}t}=-i[\omega (t),\hat h] 
\;\;, \end{eqnarray} 
with $[x,y]:=xy-yx$. Thus, the operator formalism, which is familiar in quantum theory,  
turns out to be useful in this classical context as well.  
The generator $\hat h$ is Hermitian, despite the  
dissipative character of the equations motion. 
  
The Hilbert space on which these operators act is composed of elements 
which we call {\it prequantum states}. They can be employed as usual, 
in order to calculate the observable properties of the classical system, 
which are functions ${\cal O}(\varphi ,\omega )$.   

Let us consider the evolution of those prequantum states $\psi$ which describe the 
trajectory of the classical system for an arbitrary but fixed initial condition:  
\begin{eqnarray}\label{prequ} 
\psi (\varphi ,\omega ;t)&=&
\sum_n e^{in\varphi}\psi_n (\omega ;t) 
\\ [1ex] \label{prequ1} 
&\stackrel{t\rightarrow\infty}{\longrightarrow}&
\sum_n e^{in(\varphi -\omega_it)}\psi_n (\omega_i;0)    
\;\;, \end{eqnarray} 
where $\omega_i$ is the particular fixed point to which $\omega (t)$ 
is attracted, depending on its initial condition;  
the Fourier transformation takes periodicity in the angular 
variable into account. -- Then, 
in a {\it superselection} sector where the absolutely conserved 
``quantum number'' $n$ is fixed to a particular value $n'$, the prequantum states are  
directly related to the energy eigenstates of the quantum system described by Eq.\,(\ref{G1}): 
\begin{equation}\label{psiE} 
e^{-iE_it'}\psi (E_i )
=e^{-in'\omega_it}\psi_{n'}(\omega_i;0)    
\;\;, \end{equation}  
evolving in the usual way, with $t':=n't$.
Probabilistic superpositions of prequantum states with different $\omega_i$ can be formed  
and result in pure quantum states showing interference.  

In conclusion, characteristica of quantum systems described by 
Eq.\,(\ref{G1}) can emerge from the dissipative evolution of deterministic systems beneath. 

Next, we consider quantum mechanical objects that require, for a 
complete characterization of their state,    
a set of simultaneous eigenvalues of a number of linearly independent and mutually 
commuting Hermitian operators, $\hat A_n,\;n=1,\dots ,N$, collectively denoted by $\vec A:=(\hat A_1,\dots ,\hat A_N)^t$, which are the {\it beables} \cite{BellBook}. 
These operators, with eigenvalues denoted by $A_n^j,\;j=1,\dots ,d$, 
act linearly on a finite dimensional Hilbert space, corresponding to a finite 
number of degrees of freedom. 

While a particular $\mbox{GL}(N,\mathbf{R})$ {\it symmetry of beables}~\cite{Elze08} 
has been useful 
in generalizing the previous existence theorem for the case at hand, we assume 
here for simplicity that the set of beables is fixed. 

Then, the proof of existence of a {\it deterministic model accounting for   
a finite number of beables}~\cite{Elze08} is analogous to the 
previous one, Eqs.\,(\ref{G2})--(\ref{psiE}). -- The model here 
comprises $N$ real degrees of freedom which are periodic, 
$\vec\varphi :=(\varphi_1,\dots ,\varphi_N)^t$, $\varphi_n\in [0,2\pi [$, and 
evolve according to the classical equation of motion: 
\begin{equation}\label{phivec} 
\frac{\mbox{d}\vec\varphi (t)}{\mbox{d}t}=\vec\omega 
\;\;, \end{equation} 
involving a second set of $N$ real degrees of freedom, 
$\vec\omega :=(\omega_1,\dots ,\omega_N)^t$. 

While Eq.\,(\ref{phivec}) replaces Eq.\,(\ref{G2}), 
presently the vector $\vec\omega$ evolves according to the classical equation:  
\begin{equation}\label{Fbc} 
\frac{\mbox{d}\vec\omega (t)}{\mbox{d}t}=-\kappa\frac{\partial}{\partial\vec\omega}
F^2(t)
\;\;, \end{equation} 
with $\kappa >0$. 
This equation determines $\vec\omega$, once    
its initial value $\vec\omega (t_0)$ is supplied and the function $F$ is given by: 
\begin{equation}\label{Finit} 
F(t):=\sum_{n=1}^N{\det}^2\big (\vec A-\vec\omega (t)\mathbf{1}\big )_n
\;\;, \end{equation}   
where the sum is over the components of the vector inside $(\dots )$ and 
the determinant refers to the Hilbert space on which the  
operators act that are collected in $\vec A$. 

We remark that 
operators and Hilbert space have only been introduced for convenient book keeping.   
Essentially needed, so far, are the real numbers $A_n^j$ which 
parametrize $F$, in analogy to the 
energy eigenvalues entering Eqs.\,(\ref{G3}). -- 
An important property of beables is that related eigenvalues 
are invariant under unitary transformations in Hilbert space.  
Therefore, the function $F$ had to be a scalar under such transformations. 
Other symmetry aspects are discussed in 
Ref.~\refcite{Elze08}.   

Furthermore, the sum of squares of determinants in Eq.\,(\ref{Finit})  
is zero, if and only if the $N$-dimensional vector $\vec\omega$ corresponds to  
one of the points of the $N$-dimensional finite lattice defined by  
$d\times N$ numbers $A_n^j$, i.e., by the $d$ eigenvalues of each one of the $N$ operators 
$\vec A$. 
In this way, the Eqs.\,(\ref{Fbc})--(\ref{Finit}) generalize the Eqs.\,(\ref{G3}). --    
As in the previous case, the zeros of $F$ are attractive. Thus, 
the vector $\vec\omega$ is attracted to a fixed vector with components determined 
by eigenvalues of the operators $\hat A_n$: 
\begin{equation}\label{fixp} 
\omega_n (t)\;\stackrel{t\rightarrow\infty}{\longrightarrow}\; 
A_n^{j(n)}=:\omega_n^\ast 
\;\;. \end{equation} 
Which particular eigenvalues 
contribute, indexed by $j(n),\;j=1,\dots,d$, depends on the arbitrary initial condition 
for $\vec\omega$.     

Finally, the considerations of Eqs.\,(\ref{prequ})--(\ref{psiE}) are easily generalized. 
We consider prequantum states $\psi$ which describe the trajectory 
of the deterministic system: 
\begin{equation}\label{prequF} 
\psi (\vec\varphi ,\vec\omega ;t)=
\sum_{\vec n} e^{i\vec n\cdot\vec\varphi}\psi_{\vec n} (\vec\omega ;t) 
\;\stackrel{t\rightarrow\infty}{\longrightarrow}\;
\sum_{\vec n} e^{i\vec n\cdot(\vec\varphi -\vec\omega^\ast t)}\psi_{\vec n} 
(\vec\omega^\ast;0)    
\;\;, \end{equation} 
where the fixed point $\vec\omega^\ast$ for a given initial condition and periodicity in $\vec\varphi$ are taken into account.   
All components of the vector $\vec\omega^\ast$ contribute 
to the phase of the evolving state, i.e., all $N$ Hermitian 
operators $\hat A_n$ contribute, each with one of its set of $d$ eigenvalues 
$A_n^j$. The states fall into superselection sectors that can be classified by the 
absolutely conserved vector $\vec n$. 

Three qualitatively different situations may arise. -- First, the model universe may 
find itself in a state where all components of $\vec n$ are equal, denoted by 
$\vec n'\equiv (n',\dots,n')^t$. Here, the emergent   
Hamiltonian must be identified as: 
\begin{equation}\label{Hamiltonian} 
\hat H = \sum_{n=1}^N\hat A_n 
\;\;, \end{equation}  
with an eigenvalue $E_*=\sum_{n}A_n^{j(n)}=\sum_n\omega_n^\ast$  
corresponding to a particular initial condition for the deterministic trajectory. The   
quantum states are related to the prequantum states by: 
\begin{equation}\label{psiE1}
e^{-iE_*t'}\psi (E_*)=e^{-in'\sum_n\omega_n^\ast t}
\psi_{\vec n'}(\vec\omega^\ast ;0)     
\;\;, \end{equation} 
with $t':=n't$, cf. Eq.\,(\ref{psiE}). One of the beables, corresponding 
to $\omega^\ast_1$, for example, could be eliminated in favour of the Hamiltonian 
and $E_*$, respectively, such that the above relation becomes 
$\psi (E_*)\propto \psi_{\vec n'}(E_*,\omega^\ast_2,\dots,\omega^\ast_N;0)$. 
Thus, we find degenerate energy eigenstates, which are further resolved by the 
eigenvalues of the $N-1$ remaining beables, i.e., by the values of 
$\omega^\ast_2,\dots,\omega^\ast_N$. --         
Second, assuming that all $\omega^\ast_n$ 
are of the same order of magnitude, one of the components of the superselection 
vector $\vec n$, say $n_1$, may be very much larger than all others. In this case, 
we consider the Hamiltonian: 
\begin{equation}\label{Hamiltonian1} 
\hat H_1 =\hat A_1 
\;\;, \end{equation}   
with eigenvalues $E_*=\omega_1^\ast$, which presents a valid approximation, 
as long as only sufficiently small eigenvalues $\omega^\ast_{n>1}$ have to be   
taken into account. 
In this case:
\begin{equation}\label{psiE2} 
e^{-iE_*t'}\psi (E_*)=e^{-in_1\omega_1^\ast t} 
\psi_{\vec n}(E_*,\omega^\ast_2,\dots,\omega^\ast_N;0) 
\;\;, \end{equation}  
with $t':=n_1t$. 
That is, one contribution to the phase is dominant; this leads   
to degenerate energy eigenstates, to be resolved as before. --  
There will be 
only accidental degeneracies, if any, 
in the third case, when all beables possibly contribute:    
\begin{equation}\label{Hamiltonian2} 
\hat H_{all} = \vec n\cdot\vec A 
\;\;, \end{equation}  
with eigenvalues of the form $E_*=\vec n\cdot\vec\omega^\ast$. Here, we 
obtain: 
\begin{equation}\label{psiE3} 
e^{-iE_*t}\psi (E_*)=e^{-i\vec n\cdot\vec\omega^\ast t} 
\psi_{\vec n}(\vec\omega^\ast ;0) 
\;\;. \end{equation}  
One of the eigenvalues $\omega^\ast_k$ could be replaced by 
$E_*-\sum_{m\neq k}n_m\omega^\ast_m/n_k$, 
provided $n_k\neq 0$. Thus, in this most general case, there still exist a 
unique Hamiltonian and a related energy variable, 
which govern the evolution of the emergent states. 
 
This completes our review of the existence theorems~\cite{tHooft07,Elze08}. 
They do not present constructive theorems, since spectral information 
is needed as input. In the main part of this article, we aim for a constructive theory 
instead.

\end{document}